\documentclass[twocolumn]{aastex63}

\usepackage[utf8]{inputenc}
\usepackage{newunicodechar}
\usepackage{booktabs}
\usepackage{natbib}
\usepackage{graphbox}  
\usepackage{url}
\usepackage{hyperref}
\usepackage{color}
\usepackage{rotating}
\usepackage[english]{babel}
\usepackage{breakurl}
\usepackage{graphicx,amssymb,amsmath,times,subfigure,multirow}
\usepackage{blindtext}
\usepackage{etoolbox}

\makeatletter

\usepackage{wasysym}

\begin{document}

\newcommand{\fermi}{\emph{Fermi}~}
\newcommand{\fermilat}{\emph{Fermi}~LAT~}

\newcommand{\vdag}{(v)^\dagger}
\newcommand{\aastex}{AAS\TeX}
\newcommand{\latex}{La\TeX}

\newcommand{\ltsima}{$\; \buildrel < \over \sim \;$}
\newcommand{\simlt}{\lower.5ex\hbox{\ltsima}} 
\newcommand{\gtsima}{$\; \buildrel > \over \sim \;$}
\newcommand{\simgt}{\lower.5ex\hbox{\gtsima}} 
\newcommand{\arcsecs}{\hbox{$^{\prime\prime}$}}
\newcommand{\degree}{\hbox{$^\circ$}}
\newcommand{\phflux}{photons cm$^{-2}$ s$^{-1}$}
\newcommand{\cgsflux}{erg s$^{-1}$ cm$^{-2}$}
\newcommand{\cgslum}{erg s$^{-1}$}
\newcommand{\gray}{$\gamma$-ray}
\newcommand{\grays}{$\gamma$-rays}
\newcommand{\dt}{\hbox{$\Delta$$t$}}
\newcommand{\dtr}{\hbox{$\Delta$$t_{\rm r}$}}
\newcommand{\dtg}{\hbox{$\Delta$$t_{\gamma}$}}
\newcommand{\Ms}{$M_\odot$}
\newcommand{\Ls}{$L_\odot$}
\newcommand{\Rs}{$R_\odot$}
\newcommand{\pzero}{$\pi^{\rm 0}$}
\newcommand{\Angst}{$\buildrel _{\circ} \over {\mathrm{A}}$}
\newcommand{\rninefive}{$r_{\rm 95}$}

\newcommand{\rxte}{\emph{RXTE}~}
\newcommand{\swift}{\emph{Swift}~}

\newcommand{\glm}[1]{\textcolor{red}{#1}}
\newcommand{\fede}[1]{\textcolor{red}{#1}}
\newcommand{\scp}[1]{\textcolor{blue}{#1}}

\submitjournal{The Astrophysical Journal}


\title{Fermi-LAT Discovery of a Gamma-ray Outburst from the Peculiar Compact Steep Spectrum Radiogalaxy 3C~216}
%
\shorttitle{Fermi-LAT Discovery of a Gamma-ray Outburst from CSS Radiogalaxy 3C~216}
\shortauthors{Giacchino et al.}

\author[0000-0002-0247-6884]{Federica Giacchino}
\affiliation{Istituto Nazionale di Fisica Nucleare Sezione di Roma Tor Vergata, via della Ricerca Scientifica 1, 00133, Roma, Italy}
\affiliation{ASI Space Science Data Center, via del Politecnico, 00133, Roma, Italy}

\author[0000-0001-8553-499X]{Giovanni La Mura}
\affiliation{Istituto Nazionale di Astrofisica - Osservatorio Astronomico di Cagliari, Via della Scienza 5, 09047, Selargius (CA), Italy}
\affiliation{Laborat\'orio de Instrumenta\c{c}\~ao e F\'{i}sica Experimental de Part\'{i}culas, Av. Prof. Gama Pinto 2, 1649-003 Lisboa, Portugal}

\author[0000-0001-9325-4672]{Stefano Ciprini}
\affiliation{Istituto Nazionale di Fisica Nucleare Sezione di Roma Tor Vergata, via della Ricerca Scientifica 1, 00133, Roma, Italy}
\affiliation{ASI Space Science Data Center, via del Politecnico, 00133, Roma, Italy}

\author[0000-0002-5064-9495]{Dario Gasparrini}
\affiliation{Istituto Nazionale di Fisica Nucleare Sezione di Roma Tor Vergata, via della Ricerca Scientifica 1, 00133, Roma, Italy}
\affiliation{ASI Space Science Data Center, via del Politecnico, 00133, Roma, Italy}

\author[0000-0002-8657-8852]{Marcello Giroletti}
\affiliation{INAF, Istituto di Radioastronomia, Via Gobetti 101, 40129 Bologna, Italy}

\author[0000-0001-5762-6360]{Marco Laurenti}
\affiliation{Dipartimento di Fisica, Universit\'{a} degli Studi di Roma ``Tor Vergata'', Via della Ricerca Scientifica 1, I-00133 Roma, Italy}
\affiliation{Istituto Nazionale di Fisica Nucleare Sezione di Roma Tor Vergata, via della Ricerca Scientifica 1, 00133, Roma, Italy}

\correspondingauthor{Federica Giacchino}
\email{federica.giacchino@roma2.infn.it}

\begin{abstract}

3C~216 is an extra-galactic radio source classified as a compact steep spectrum (CSS) object, associated with the source 4FGL~J0910.0+4257 detected by the Large Area Telescope (LAT) on board the \textit{Fermi} Gamma-ray Space Telescope. The source exhibits extended radio structures as well as an inner relativistic jet. In general, jets accelerated by active galactic nuclei (AGNs) are efficient sources of non-thermal radiation, spanning from the radio band to X-ray and gamma-ray energies. Due to relativistic beaming, much of this radiation, particularly in the high-energy domain, is concentrated within a narrow cone aligned with the jet's direction. Consequently, high-energy emission is more easily detected in \textit{blazars}, where the jet is closely aligned with the line of sight of the observer. Beginning in November 2022, 
\textit{Fermi}-LAT observed increased gamma-ray activity from 3C~216, culminating in a strong outburst in May 2023. This event was followed up by observations from the \textit{Neil Gehrels Swift Observatory} telescope. In this work, we perform a careful analysis of the multifrequency data (gamma-ray, X-ray, UV, optical) collected during this observational campaign. We find that the spectral energy distribution of the flaring source evolves in a coherent way, supporting a common origin for the multifrequency emission. These results suggest that the spectral energy distribution (SED) observed during the outburst was dominated by a single emission zone, where synchrotron self-Compton processes (SSC) played a primary role. Since single-zone SSC models have typically less free parameters than multi-zone alternatives, they are a powerful probe of the physical conditions of the high-energy emitting regions. Therefore, observing SSC radiation even in CSS sources improves our understanding of the production of high-energy radiation in AGN jets.

\end{abstract}


  \section{Introduction} \label{intro}


The active galactic nucleus (AGN) 3C~216 ($z=0.670$, \citealt{1980ApJ...236..419S}) is classified as an extragalactic CSS radio source. On arcsecond scales, it consists of a central component surrounded by a more extended halo structure with an angular size of $4.5''$, 
corresponding to a projected linear size (LS) of $56$\,kpc \citep{2021MNRAS.507.4564P}. Although its radio spectrum peaks at low frequency ($\nu < 0.5$\,GHz), an upturn is observed at a few GHz, attributed to the presence of a central flat-spectrum core plus a faint halo component \citep{1995AJ....110..522T}. The core exhibits a significant misalignment with the outer structure, along with superluminal jet component motion, with a velocity of approximately $\sim 4c$ \citep{1993A&A...271...65V,2000PASJ...52..983P}. While the compact nature of CSS sources is typically attributed to their young age within an evolutionary framework \citep{1995A&A...302..317F}, the characteristics of the central component, a pronounced optical polarization and its variability \citep{1991ApJ...375...46I} strongly suggest the presence of a blazar core.

The Large Area Telescope \citep[LAT,][]{Atwood09} onboard the \textit{Fermi Gamma-ray Space Telescope}, is a pair-conversion telescope designed to detect photons in the energy range from $20$ MeV to $2$ TeV. The LAT observed a pronounced enhancement in the gamma-ray activity from the direction of the source on 2022-12-08~\citep{2022ATel15801....1L} and on 2023-05-01 \citep{2023ATel16024....1G}. Multifrequency observations of the source  were carried out,  thanks to a \textit{Swift} Target of Opportunity (ToO) request, in order to associate the $\gamma$-ray flare with activity at different wavelengths. The observations across various bands provide insights into both the origin of the $\gamma$-ray emission and the structure of the relativistic jet. During the flare, the source exhibited a hard gamma-ray spectrum, as opposed to the typically soft one observed in quiescent periods, together with coherent variability from low energies up to $\gamma$ rays. This behavior can be explained in terms of a single-zone emission model, where it is likely that synchrotron self-Compton radiation dominated the emission across all frequencies during the outburst, with evidence of a cooling process in the following days.

The paper is structured as follows: in Sec.~\ref{fermilatobservation} we present the detection of the exceptional gamma-ray activity observed by \textit{Fermi}-LAT in May 2023, with a detailed analysis of the \textit{Fermi}-LAT data and its comparison with long term observations given in Sec.~\ref{fermilatanalysis}; in Sec.~\ref{swiftobservations} we describe the \textit{Swift} ToO observations of 3C~216, detailing their analysis in Sec.~\ref{swiftanalysis}; in Sec.~\ref{sec:discussion} we discuss our results and, finally, in Sec.~\ref{sec:conclusions} we summarize our conclusions.

\section{Fermi-LAT observations} \label{fermilatobservation}
3C 216 is located at $R.A. = 137.38957\,$deg, $DEC. = 42.89624\,$deg \citep[J2000,][]{2005AJ....129.1163P}. It is associated with the $\gamma$-ray source 4FGL~J0910.0+4257, with coordinates $R.A. = (137.51 \pm 0.11)\,$deg, $DEC. = (42.96 \pm 0.11)\,$deg in the third release of the 4FGL catalog \citep[4FGL-DR3,][]{2022ApJS..260...53A}. In the course of its regular monitoring of the sky, \textit{Fermi}-LAT detected an increase of activity from the direction of this source starting from the mid of November 2022. On 2023-05-01, a quick-look analysis showed that the source exhibited a sudden outburst, achieving a daily averaged $\gamma$-ray flux of $\langle \Phi \rangle_{daily} = (1.32 \pm 0.15) \times 10^{-6}\, \mathrm{ph\, cm^{-2}\, s^{-1}}$, a factor 176 times higher than the average flux $\langle \Phi \rangle_\gamma = (7.5 \pm 1.4) \times 10^{-9}\, \mathrm{ph\, cm^{-2}\, s^{-1}}$ reported in the 4FGL-DR3 catalog over the $0.1-100\,$GeV energy range. The photon index decreased from the catalog value of $(2.52 \pm 0.10)$ to $(2.11 \pm 0.09)$. This event represented both the highest daily flux and the hardest spectral state reported for the source to date \citep{2023ATel16024....1G}.

Due to the exceptional magnitude of the event, multi-frequency follow-up observations were requested by activating a \textit{Swift} ToO, which detected 3C 216 in a flaring state in optical, UV, and X-rays, confirming its association with 4FGL~J0910.0+4257. We therefore analyzed the $\gamma$-ray history of the source and its associated X-ray and optical/UV observations to characterize the origin of its flaring activity.

\begin{figure}[h!]
  \centering
  \includegraphics[width=0.98\linewidth]{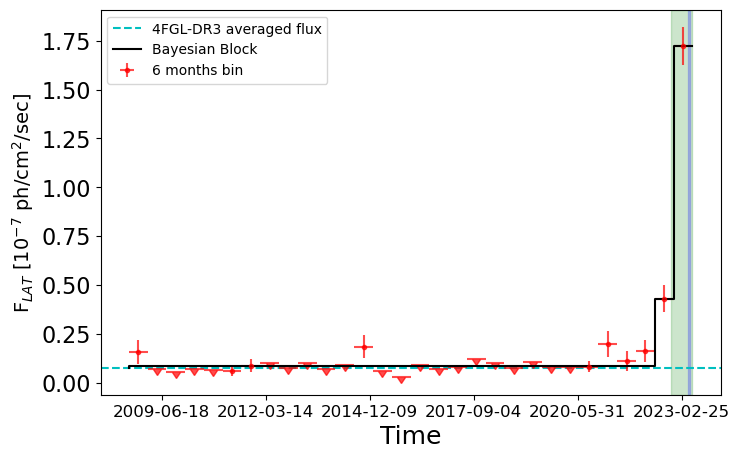}
  \includegraphics[width=0.98\linewidth]{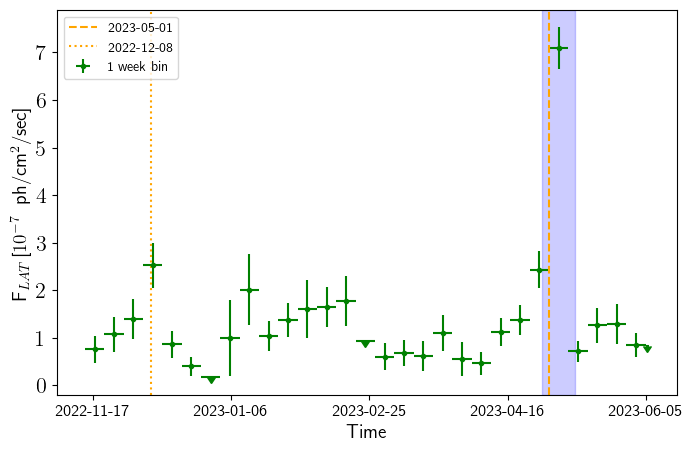}
  \includegraphics[width=0.98\linewidth]{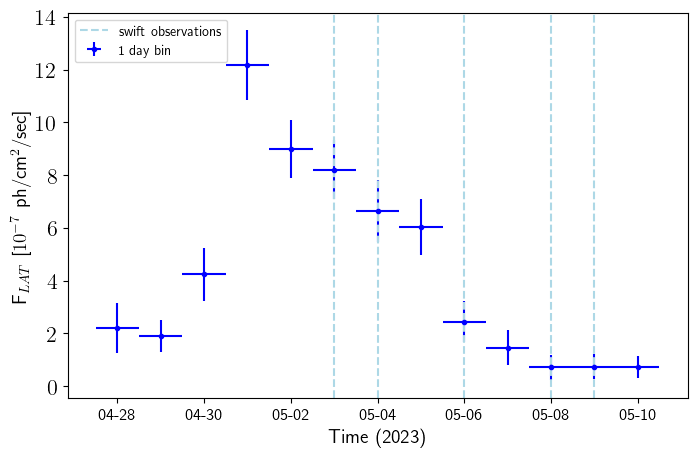}
  \caption{\textit{Fermi}-LAT light curves of 3C~216 for the three periods considered here with different binning: in the top panel \ (\emph{red}) the total period with $6$-month binning; in the middle panel (\emph{green}) the flare period in weekly binning; in the bottom panel (\emph{blue}) the flare peak in daily binning. In the top panel the dashed cyan line represents the 4FGL-DR3 averaged flux, and the plain black bins are computed by the Bayesian Block algorithm~\citep{2013ApJ...764..167S}. The shaded areas of each panel corresponds to the period detailed in the following one. The dashed light blue lines in the bottom panel mark the days where \textit{Swift} observations were obtained. F$_{\textit{LAT}}$ is the flux detected by LAT from $100$ MeV to $300$ GeV. Upper limits are reported when $TS\leq 10$.} \label{fig:lc}
\end{figure}
\subsection{Gamma-ray analysis}\label{fermilatanalysis}
In order to characterize the status of 3C~216 during the outburst and to compare it with its long-term averaged properties, we analyzed the data collected by \textit{Fermi}-LAT over three distinct periods. The first, which we refer to as the \textit{total monitoring period}, covers all the \textit{Fermi}-LAT observations of the source, from the start of science operations on 2008-08-04 up to 2023-06-06, approximately one month after the main $\gamma$-ray outburst. The second, which we call the \textit{flare period}, corresponds to the start of the weekly rising trend in the activity of the source, on 2022-11-14, up to the end of the total period. The third one, which we call the \textit{flare peak}, extends from 2023-04-28 to 2023-05-10, covering the time range corresponding to the highest level of daily $\gamma$-ray activity and the subsequent \textit{Swift} follow-up campaign. For all these periods we performed a light curve analysis, plotted in Fig.~\ref{fig:lc}, and a standard binned analysis.

\begin{table}[t]
\begin{center}
\caption{Table of \textit{fermipy} analysis parameters. \label{tab:analysis}}
\begin{tabular*}{\linewidth}{l@{\extracolsep{\fill}}c}
\hline
\hline
Parameter name & Value \\
\hline
Time domain (Gregorian) & $2008/08/04$ to  $2023/06/06$\\
Energy range & 100 MeV to 300 GeV \\
IRF & \texttt{P8R3\_SOURCE\_V3}\\
Event Type & FRONT + BACK \\
Point Source Catalog & 4FGL-DR3 \\
ROI size & $15^{\circ} \times 15^{\circ}$ \\
Pixel size & $0.1^{\circ}$ \\
Bins per energy decade & 8 \\
Galactic diffuse model & \texttt{gll\_iem\_v07.fits}\\
Isotropic diffuse model &\texttt{P8R3\_SOURCE\_V3\_v1.txt}\\
\hline
\end{tabular*}
\end{center}
\end{table}
In our study, we used \textit{fermipy} \texttt{v1.2.0}\footnote{\url{https://fermipy.readthedocs.io/en/latest/}} \citep{2017ICRC...35..824W} and the \textit{FermiTools} \texttt{v2.2.0}.\footnote{\url{https://fermi.gsfc.nasa.gov/ssc/data/analysis/software/}} We selected all the observations in the $[0.1-300]\,$GeV energy range, using Pass 8 events~\citep{2013arXiv1303.3514A} and all the available photons of the SOURCE class, excluding those arriving with zenith angles greater than $90^{\circ}$ for energies smaller than $1$ GeV, otherwise greater than $105^{\circ}$. Moreover, we chose a region of interest (ROI) with an aperture radius of $15^{\circ}$ around the target, with a pixel size of $0.1^{\circ}$ and $8$ evenly spaced logarithmic energy bins. We utilized the \texttt{P8R3\_SOURCE\_V3} instrumental response functions (IRFs), along with the galactic diffuse model \texttt{gll\_iem\_v07.fits} and the isotropic diffuse model \texttt{P8R3\_SOURCE\_V3\_v1.txt}. The model used for the analysis includes all sources in the 4FGL-DR3 catalog\footnote{\url{https://fermi.gsfc.nasa.gov/ssc/data/access/lat/BackgroundModels.html}} located at a distance $\leq 20^{\circ}$ from 4FGL~J0910.0+4257. A summary of the analysis parameters is reported in Table~\ref{tab:analysis}.

\begin{figure}[t]
    \centering
\includegraphics[width=1\linewidth]{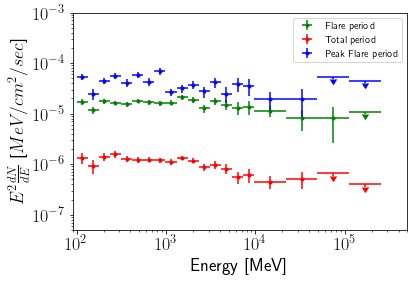}
    \caption{SED of 3C~216 for three periods: in \emph{flare peak} (blue), in \emph{flare} (green), and \emph{total} (red) periods. Upper limits are reported when $TS \leq 10$. \label{plot:SED}}
\end{figure}
For the spectral analysis we left the model parameters of the isotropic and diffuse background as well as the sources within $3^{\circ}$ of our target free to vary. To produce the spectral energy distribution (SED) of all the three periods, we used eight logarithmically spaced bins between $100$ MeV and $10$ GeV and four energy bins from $10$ GeV and $300$ GeV, obtaining the plot in Fig.~\ref{plot:SED}. The result clearly shows that the \textit{flare peak} (blue) and the \textit{flare} (green) period fluxes are much higher than the values obtained considering all the \textit{Fermi}-LAT observations (red), confirming an interesting enhanced activity. For the light curve analysis shown in Fig.~\ref{fig:lc}, instead, we allowed the flux normalization of the sources within $3^{\circ}$ of the target to vary freely while freezing all the other parameters from the baseline analysis. The top panel displays the \emph{total period} with $6$-month time bins, the middle panel shows the \emph{flare period} in a weekly binning, and the botton panel presents the evolution leading to the \emph{flare peak} and the subsequent follow-up, using $1$-day time bins from $2023$-$04$-$28$ to $2023$-$05$-$10$. For the total period we have identified statistically
significant variations compared to the 4FGL-DR3 averaged flux with the Bayesian Blocks algorithm \citep{2013ApJ...764..167S} as implemented in astropy\footnote{\url{https://docs.astropy.org/en/stable/api/astropy.stats.bayesian_blocks.html}}.

The analysis of the \emph{flare period} reveals that the flare activity achieved a first maximum on 2022-12-8, marked by the orange dotted line in the middle panel of Fig.~\ref{fig:lc},followed by an epoch of moderate activity before the exceptional outburst occurring on 2023-05-01, highlighted by the orange dashed line. The blue curve in the \emph{peak} period illustrates the sharp rise of the flux leading to the flare peak (bottom panel of Fig.~\ref{fig:lc}),  while the light blue dashed lines indicate the \textit{Swift} observations conducted during the follow-up campaign. The maximum flux obtained by the analysis is $\Phi_{max}=(1.22\pm 0.13)\times 10^{-6}$ ph cm$^{-2}$ s$^{-1}$ with a photon index of $2.06\pm0.12$. The shaded areas in the top and middle panels refer to the time range of the following ones.

\begin{table*}[t]
    \caption{List of averaged spectral parameters for three periods (flare peak, flare, total): photon index for power-law $\Gamma$, normalization factor for power-law $\Phi_0$, slope of the log-parabola $\alpha$, curvature log-parabola $\beta$, normalization factor for log-parabola $N_0$. We include, for reference, also the values reported in the 4FGL-DR4 catalog \citep{2023arXiv230712546B}}. \label{tab:SpectralPars}
    \begin{tabular}{cccccc}
      \hline
      \hline
      Period & $\Gamma$ & $\Phi_0$ $[\mathrm{MeV^{-1}\, cm^{-2}\, s^{-1}}]$ & $\alpha$ & $\beta$ & $N_0$ $[\mathrm{MeV^{-1}\, cm^{-2}\, s^{-1}}]$ \\
      \hline
      \emph{Flare Peak} & $1.97\pm 0.05$ & $(5.87\pm 0.46)\times 10^{-11}$ & $1.79\pm 0.23$& $0.12\pm 0.09$ & $(3.83\pm 0.72)\times 10^{-11}$\\
      \emph{Flare} & $2.03 \pm 0.01$ & $(2.45\pm 0.09)\times 10^{-11}$&$1.99\pm 0.16$ & $0.06 \pm 0.03$ & $(1.73\pm 0.25)\times 10^{-11}$\\
      \emph{Total} & $2.17 \pm 0.04$ & $(1.69\pm 0.07)\times 10^{-12}$ &$1.95 \pm 0.20$ & $0.12 \pm 0.07$ & $(1.02\pm 0.47)\times 10^{-12}$\\
      4FGL-DR4 & $2.43\pm 0.07$ & $(8.05\pm 0.71)\times 10^{-13}$ &$2.24\pm 0.15$ & $0.17\pm 0.10$ & $(9.02\pm 0.91)\times 10^{-13}$\\
      \hline
    \end{tabular}
\end{table*}
\subsection{$\gamma$-ray spectral properties}
\begin{figure}[t]
  \centering
  \includegraphics[width=1\linewidth]{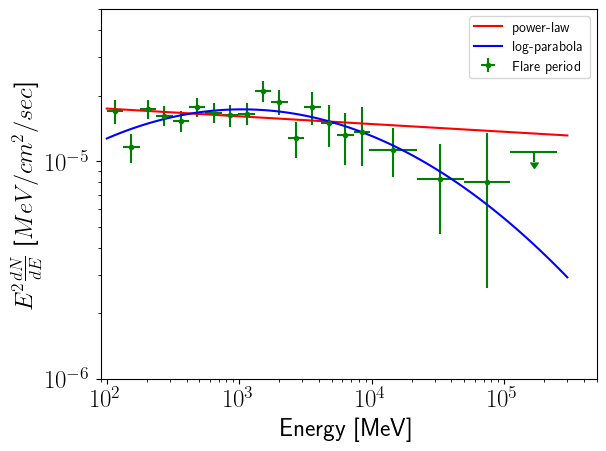}
  \caption{SED of 3C~216 during the flare period. The red line is the power-law function and the blue line is the log-parabola function whose spectral parameters are described in Table~\ref{tab:SpectralPars}. The upper limit is reported when $TS\leq 10$. \label{plot:spectrumfit}}
\end{figure}

AGN gamma-ray spectra are generally well represented by either a log-parabola (LP) function or by a power-law (PL) one. The LP function is defined as:

\begin{equation}
    \frac{dN}{dE}=\Phi_0\Big(\frac{E}{E_0}\Big)^{-\alpha-\beta Log(E/E_0)}
\end{equation}

\noindent where the normalization factor $\Phi_0$ (MeV$^{-1}$ cm$^{-2}$ sec$^{-1}$) is the flux density at $E_0$, $\alpha$ is the spectral index, $\beta$ is the curvature, and $E_0$ (MeV) is the scale parameter. The PL function, instead, is defined as:

\begin{equation}
\frac{dN}{dE}= N_0\,\Big(\frac{E}{E_b}\Big)^{-\Gamma}
\end{equation}

\noindent where the parameters are the normalization factor $N_0$ in ($\mathrm{MeV^{-1}\, cm^{-2}\, s^{-1}}$), the spectral index $\Gamma$ and the energy scale $E_b$ (MeV). According to ~\citet{2022ApJS..260...53A}, the sources are represented with a curved spectral model when $TS_{curv} = 2[log\mathcal{L}_{curv} - log\mathcal{L}_{PL}] > 4\, ( = 2\sigma)$. In our case the likelihood for curvature is evaluated by testing LP against PL spectral models. All the three periods are characterized by a larger $TS_{curv}$ than threshold: $TS_{LP}= 3.0\sigma$\ for the \emph{total period}, $TS_{LP}= 3.2\sigma$\ for the \emph{flare period}, and $TS_{LP}= 3.1\sigma$\ for the \emph{flare peak} period. Notably, the 4FGL-DR3 catalog reports a curvature significance of $TS_{LP}\sim 1.68 \sigma$ for this source and the preferred spectral model is a power-law. On the contrary, the longer monitoring of 4FGL-DR4~\citep{2023arXiv230712546B} points to a curvature significance of $TS_{LP}\sim 2.39\sigma$, supporting a curved spectrum, in agreement with our result. Likely the flaring activity increased the number of photons collected from this source allowing a more accurate reconstruction of its spectral form. As a consequence, the spectrum during the flaring period is well represented by a log-parabola model, as shown in Fig.~\ref{plot:spectrumfit}.

The evolution of the spectral state observed for the total period, the flare period and the flare peak period can be appreciated by looking at the corresponding spectral index and curvature parameters, listed in Table~\ref{tab:SpectralPars}, in comparison with the values reported in 4FGL-DR4 \citep{2023arXiv230712546B}. We see from these results that the \emph{flare period} is characterized by a spectral hardening, with respect to the average state of the source, that is detected as a decrease of the power-law index and of the log-parabola slope, becoming more evident at the time of the peak flux.
 
\begin{figure}[t]
  \begin{center}
    \includegraphics[width=1\linewidth]{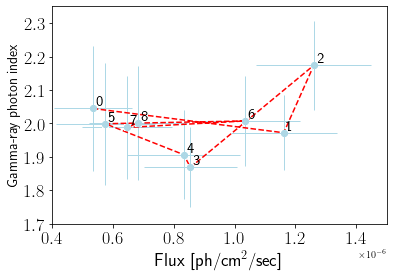}
    \includegraphics[width=1\linewidth]{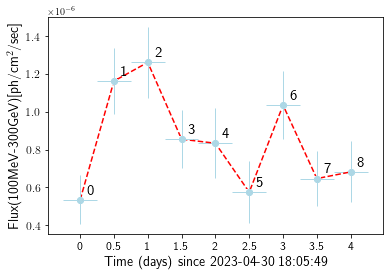}
    \caption{Top panel: the gamma-ray photon index of 3C~216 as a function of the gamma-ray flux.  Bottom panel: the light curve of  3C~216. In both panels, the numbers indicate the chronological order of  the observations, extracted as $12\,$hr bins starting from $2023$-$04$-$30$ at $18$:$05$:$49$ (designed as number $0$) and finishing point on $2023$-$05$-$04$ at $18$:$05$:$49$ (designed as number $8$).} \label{plot:index}
  \end{center}
\end{figure}

To further characterize the spectral variability during the flare peak, we studied the spectral variability in relation to the flux, performing a $12\,$hr binned analysis from 2023-04-30 to 2023-05-03 (Fig.~\ref{plot:index}). In this short time range, the spectral curvature is not particularly significant, and we can use a PL model to obtain a more accurate evaluation of the spectral variability. This analysis aims to characterize the spectral behaviour during the main $\gamma$-ray outburst and the subsequent decreasing stage. The observations are numbered to represent their chronological order. In spite of the rather large uncertainties, the evolution shown in the upper panel is suggestive of an anticlockwise trail for the first four bins (numbered from 0 to 3), while the spectral index appears to become more stable at later times. This is consistent with the expectations implied by a single-zone SSC scenario, where the radiative efficiency and the energy distribution of the emitting particles are subject to the competing effects of a fast energy injection, occurring nearly simultaneously for all particles, and a radiative cooling that affects high energy particles more quickly than low energy ones ~\citep{1998A&A...333..452K}. In the bottom panel of Fig.~\ref{plot:index}, we associated the light curve with the corresponding spectral data points to illustrate the flux variability over time.

\section{{\it Neil Gehrels Swift} XRT and UVOT data}\label{swiftobservations}

Following the powerful gamma-ray outburst of 3C~216, detected on 2023-05-01 \citep{2023ATel16024....1G}, a Target of Opportunity (ToO) request was sent to the \textit{Swift} Gamma-Ray Burst Mission \citep{2004ApJ...611.1005G}. \textit{Swift} executed five visits of the target, on 2023-05-03, 04, 06, 08 and 09, accumulating observations with both the XRT and UVOT instruments. The observations were executed 
with a roughly regular spacing of $1.5$ days between subsequent visits.
During the observations, the XRT instrument worked in photon counting mode, while UVOT performed a sequence of exposures using the $V$, $B$, $U$, $UW1$, $UM2$ and $UW2$ photometric filters.

3C~216 was detected in all the X-ray/UV/optical bands with values of the X-ray flux of the order of $\Phi_{X}\sim (2.3 - 3.2) \times 10^{-12}\,\mathrm{erg/cm^2/s}$, UV magnitudes in the range $14.6$-$15.8$, and optical ones in the range $15.2$-$16.5$. The results of these observations, concerning the UV, optical and X-ray fluxes, are detailed in Table~\ref{tab:ParametersX}, together with the X-ray photon indices and a comparison with a previous set of observations carried out on 2010-10-21.

\begin{table*}
    \caption{Table of fluxes observed by \textit{Swift} for UVOT (in six bands) and XRT (from $0.3$ to $10$ keV). Photon index $\Gamma_X$ for spectral model in X-ray. The error are $1\sigma$. Fluxes are in units of $10^{-12}\, \mathrm{erg\, cm^{-2}\, s^{-1}}$}. \label{tab:ParametersX}
    \begin{tabular}{ccccccc}
      \hline
      \hline
      Time & 2010-11-21 & 2023-05-03 & 2023-05-04 & 2023-05-06 & 2023-05-08 & 2023/05/09 \\
      \hline
      UVOT UVV & $0.06 \pm 0.01$ & $12.87 \pm 0.18$ & $11.97 \pm 0.22$ &$9.18 \pm 0.17$ &$6.00\pm0.15$ &$4.31\pm0.11$ \\
      UVOT UBB & $0.08 \pm 0.01$ &$14.62 \pm 0.25$ & $14.67 \pm 0.34$ & $11.00 \pm 0.26$ & $6.84 \pm 0.22$ & $4.97 \pm 0.16$ \\
      UVOT UUU & $0.05 \pm 0.01$ & $14.78 \pm 0.23$ & $14.12 \pm 0.32$ & $11.18 \pm 0.23$ & $6.70 \pm 0.22$ & $5.27 \pm 0.16$ \\
      UVOT UM2 & $0.01 \pm 0.01$ & $16.54 \pm 0.28$ & $15.96 \pm 0.27$ & $13.29 \pm 0.25$ & $8.87 \pm 0.26$ & $6.31 \pm 0.16$ \\
      UVOT UW1 & $0.02 \pm 0.01$ & $17.46 \pm 0.31$ & $17.40 \pm 0.31$ & $13.48 \pm 0.29$ & $9.45 \pm 0.28$ & $7.18 \pm 0.19$\\
      UVOT UW2 & $0.02 \pm 0.01$ & $16.37 \pm 0.39$ & $15.66 \pm 0.46$ & $14.34 \pm 0.44$ & $9.30 \pm 0.40$ & $7.21 \pm 0.29$ \\
      \hline
      XRT & $1.54 \pm 0.41$ & $3.24 \pm 0.42$ & $2.36 \pm 0.42$ & $2.40 \pm 0.49$ & $2.26 \pm 0.61$ & $2.84 \pm 0.72$ \\
      \hline
      $\Gamma_X$ & $1.24 \pm 0.57$ & $1.91 \pm 0.11$ & $2.03\pm 0.16$ & $1.58\pm 0.17$ & $1.62\pm 0.23$ & $1.29\pm 0.18$ \\
      \hline
    \end{tabular}
\end{table*}
\subsection{Analysis of \textit{Swift} data}
\label{swiftanalysis}
The UVOT data were processed according to a standard UVOT software analysis, using \texttt{heasoft-6.32.1}\footnote{\url{https://heasarc.gsfc.nasa.gov/docs/software/heasoft/}}. At first, we combined the exposures of each filter with \texttt{uvotimsum}, so to obtain one image per visit per filter. Then we extracted the flux and magnitude in each band-pass, together with their associated errors, using the task \texttt{uvotdetect}. 3C~216 was clearly detected in a high state, achieving a statistical significance larger than $20\sigma$\ in all the visits and with all the photometric filters. The first \textit{Swift} observation recorded a $U$-band magnitude of $(14.67 \pm 0.02)\,$mag, more than 3 magnitudes brighter than the archival value of $17.88\,$mag reported in literature in the same band \citep{Ryle64}, while subsequent observations traced a steadily decreasing trend. In spite of the long time elapsed since this archival photometric determination, the source has also been observed more recently in \textit{ugriz} photometry by the Sloan Digital Sky Survey, resulting in even fainter magnitude values in all pass-bands \citep{Adelman08}, and by \textit{Swift} itself in 2010, obtaining magnitudes of the order of $19$. We can therefore conclude that the UVOT magnitudes observed during the gamma-ray outburst correspond to a high state of the source. In order to recover the intrinsic state, we corrected the observed fluxes, accounting for the effects of foreground extinction due to the Milky Way's interstellar medium. Since 3C~216 is located far away from the Galactic Plane, it is subject to a modest reddening effect, having $A_V = 0.052$\,mag, $A_B = 0.069$\,mag and $A_U = 0.082$\,mag \citep{Schlafly_2011}. To correct the $UV$ pass-bands, we derived the extinction coefficients for the $UW2$, $UM2$ and $UW1$ according to \citet{Yi_2023}.

The XRT data were processed using \texttt{xrtpipeline v3.7.0}\footnote{\url{https://heasarc.gsfc.nasa.gov/lheasoft/ftools/headas/xrt.html}}. For each observation, spectral analysis of the source was performed within a $35$ arcsec radius from the target, with background emission subtracted from an annular region between an inner radius $r_{in}\sim100$ arcsec and an outer radius $r_{out}\sim400$ arcsec, centered on the position of 3C~216. The spectra were then binned to ensure a minimum of one count per bin and modeled using the \texttt{XSPEC v12.13.1e} \citep{1999ascl.soft10005A} package, employing the Cash statistic for minimization~\cite{1979ApJ...228..939C}. Fluxes were extracted in the soft ($[0.5-2]$ keV) and hard ($[2-10]$ keV) bands. The adopted model is a power-law modified by neutral Galactic absorption ($TB_{abs} \times z_{pow}$).

\begin{figure*}[t]
  \begin{center}
    \includegraphics[width=1\linewidth]{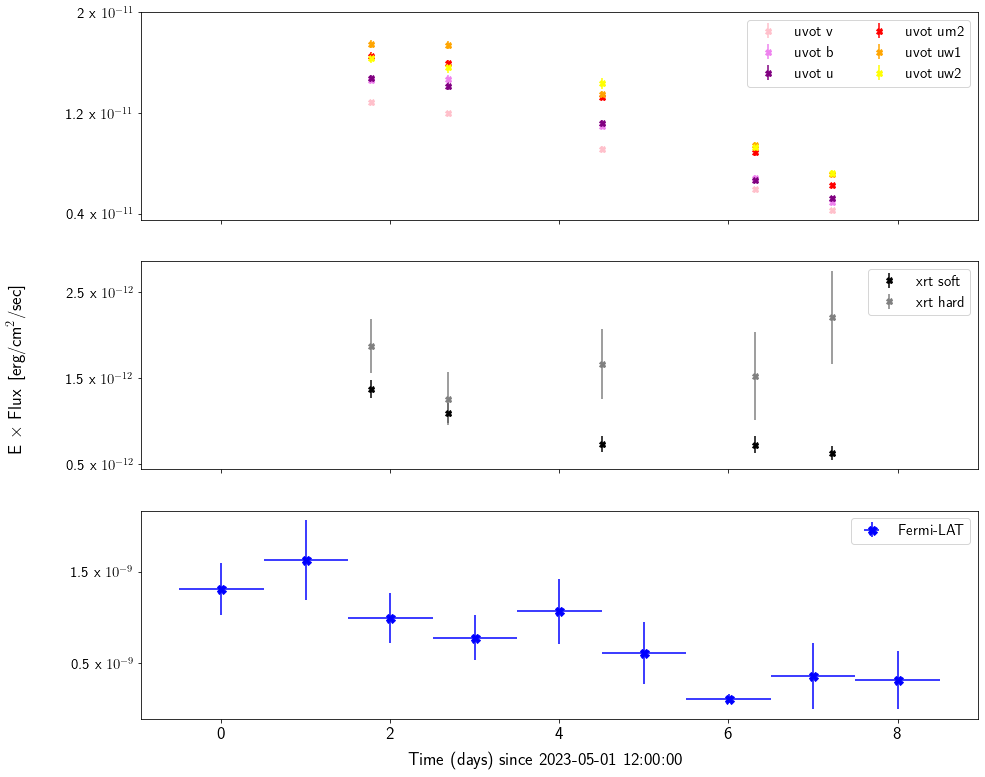}
    \caption{Light curves from UVOT data across six filters (\emph{top panel}), XRT data in two energy ranges (\emph{medium panel}), and \textit{Fermi}-LAT daily binned data (\emph{bottom panel}) collected during the \textit{Swift} follow-up campaign}. \label{plot:totallc}
  \end{center}
\end{figure*}
In Fig.~\ref{plot:totallc} we present the light curves across different frequency bands for the period covered by the \textit{Swift} observations: the UVOT data for the six filters are shown in the top panel, the XRT data for the soft and hard bands in the middle panel, and the LAT data in the bottom panel. The observed trend indicates a stable decreasing evolution in the light curves across all spectral bands, with the exception of the hard X-ray domain, which appears to be roughly constant or to exhibit a slight increase in activity. This behavior is consistent with the expectations of coherent emission from a single zone model. In this case, indeed, we would expect the energetic particles present in the jet to produce initially strong synchrotron radiation up to the UV and soft X-ray domain, while effectively scattering low energy photons up to the $\gamma$-ray domain through inverse Compton processes. However, as time passes and the radiating particles lose energy, the synchrotron radiation would gradually fade off and the inverse Compton scattering would rather produce hard X-ray radiation, instead of $\gamma$-rays.

\begin{figure*}[t]
  \begin{center}
    \includegraphics[width=0.45\textwidth]{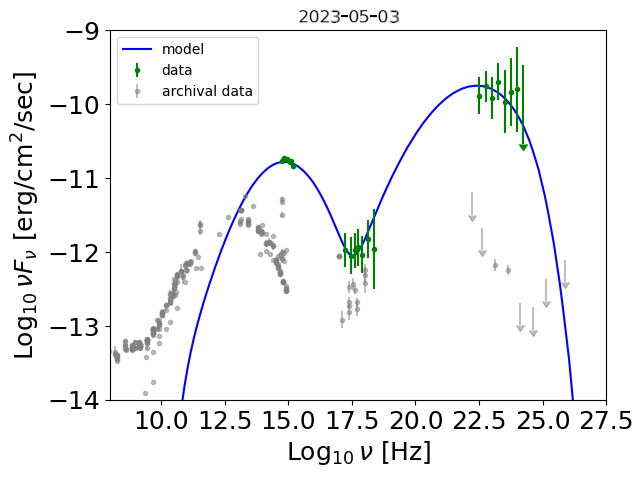}
    \includegraphics[width=0.45\textwidth]{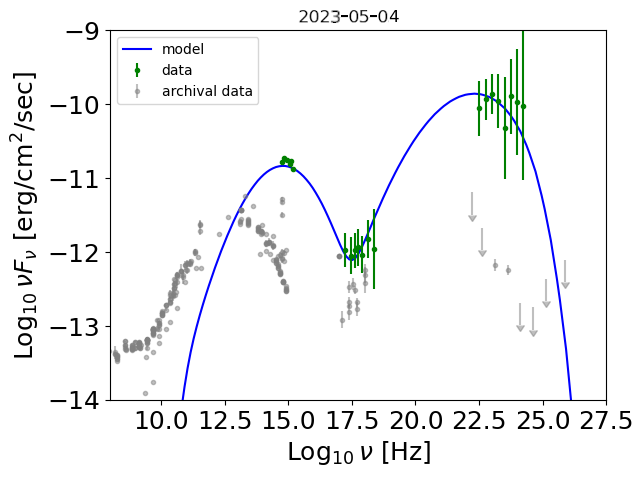}
    \includegraphics[width=0.45\textwidth]{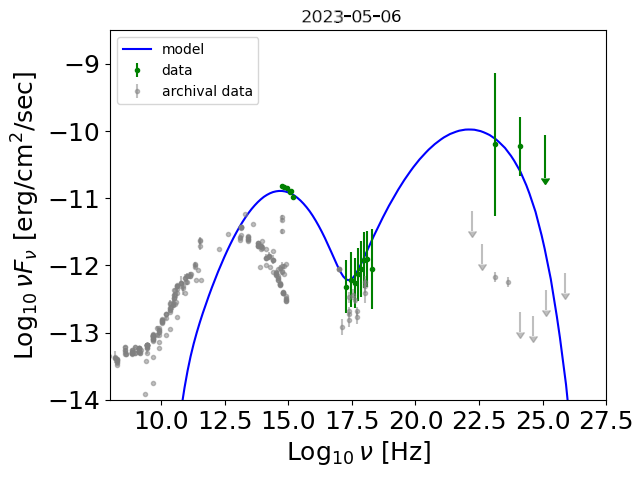}
    \includegraphics[width=0.45\textwidth]{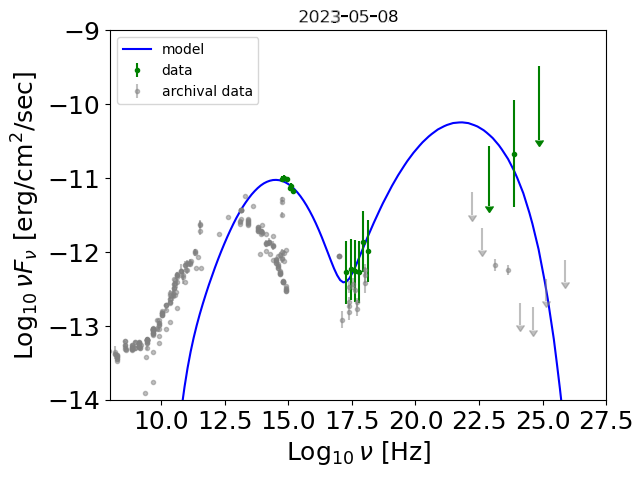}
    \includegraphics[width=0.45\textwidth]{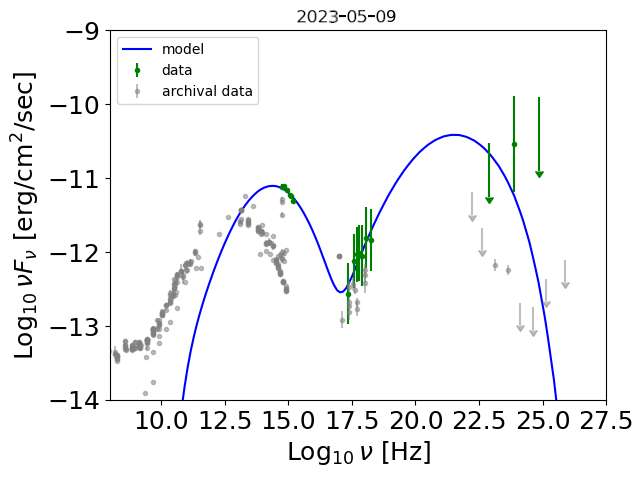}
  \end{center}
\caption{SED for the broadband emission of 3C~216, spanning from UV to gamma-ray, based on five datasets observed during the flare peak period. The blue line represents the SSC model emission computed with JetSeT. The green dots indicate the data collected by UVOT, XRT, and \textit{Fermi}-LAT, along with their respective uncertainties. Upper limits are reported when $TS\leq 10$. Light gray points in the background represent archival, non-simultaneous data from literature, specifically from the following sources: 3C, NRAO, Ohio \citep{1970ApJS...20....1D}, FIRST at VLA, Green Bank GB6 and North surveys, combined NRAO and Parkes survey, NVSS, VLBA, Planck  ERCSC, PCCS1F and PCCS2F catalogs, WMAP catalog, TWOMASS and WISE infrared catalogs, USNO and SDSS2,6,7, optical surveys, Einstein and ROSAT X-ray source catalogs, \textit{Swift} XRT SWXRT1 and 1SXPS catalogs \citep{2013A&A...551A.142D, 2014ApJS..210....8E}, Chandra ACIS source catalog \citep{2015ApJS..220....5M}, and the Fermi-LAT DR3 source catalog \citep{2022ApJS..260...53A} data. These data were extracted from the SSDC SED Builder and NED archives.} \label{fig:sedmulti}
\end{figure*}

\section{Discussion}
\label{sec:discussion}
The regular $\gamma$-ray monitoring of 3C~216, shown by Fig.~\ref{fig:lc} suggests that the source entered a high activity state extending over several months. However, the outburst that resulted in the highest observed daily flux developed in a time scale of approximately 2 days and it is therefore likely to originate from a region of limited size. The presence of multi-frequency follow-up observations allows us to obtain a series of snap-shots of the status of the SED in the days immediately after the $\gamma$-ray record state observed on 2023-05-01 and, subsequently, to attempt an interpretation of its origin.

To model the radiative emission, we used \textit{JetSet} \citep{2020ascl.soft09001T}, an open source C/Python tool for reproducing the radiative and acceleration processes acting in relativistic jets and Galactic objects (beamed and unbeamed), allowing numerical models to be fit to the observed data. \textit{JetSet} provides models to calculate SEDs obtained by combining synchrotron-self Compton and external Compton (EC), using different types of source seed radiation field, as well as the Cosmic Microwave Background (CMB). We decided to analyze the time variability of the multi-frequency SED to understand the physical process of the flare activity of 3C~216. We chose to construct our base-line model with the data set corresponding to 2023-05-03, because it is the earliest time for which we obtained simultaneous multi-frequency coverage of the source. In principle, there are many possibilities that could explain the observed $\gamma$-ray emission, including hadronic and leptonic processes, contributions from external radiation fields and multi-zone emission scenarios. However, such models are characterized by a rather large number of free parameters and they require rather complicated fine tuning to predict a coherent SED evolution, particularly in the case of multi-zone scenarios. If, on the other hand, a single-zone model is to be preferred, it is very likely that the $\gamma$-ray production mechanism is IC scattering by charged relativistic particles, which will also produce strong synchrotron emission. We therefore decided to reproduce the first available multi-frequency SED in the framework of a SSC model \citep{1974ApJ...188..353J}, testing whether the SEDs observed at later times could be interpreted by the same model, after accounting for energy loss, as it would be expected in this case. Another advantage of the SSC scenario over other possible alternatives is the limited number of model parameters, which can be reduced to $10$ in the case of a log-parabola particle energy distribution.

\begin{table}[t]
\caption{Geometric and magnetic properties of the emitting region (fixed model parameters).}
\begin{tabular*}{\linewidth}{l@{\extracolsep{\fill}}c}
\hline
\hline
Parameter & Value\\
\hline
Size of the spherical emitting region & $R = 8.3 \times 10^{15}\,$cm \\
Particle density & $n = 10^3\,\mathrm{cm^{-3}}$ \\
Intensity of the magnetic field in $R$ & $B = 1\,$G \\
Bulk Lorentz factor in $R$ & $\Gamma = 8.5$ \\
Jet viewing angle & $\theta \simeq 1/\Gamma = 4^{\circ}$ \\
\hline
\end{tabular*}
\label{tab:jetpar}
\end{table}
Our strategy consisted in dividing the model parameter in a set of $5$ fixed values, listed in Table~\ref{tab:jetpar} and used to describe the size of the radiating zone, its magnetic field and its bulk motion, and another set of variable values, listed in Table~\ref{tab:ModelPars}, which account for the energy evolution of the radiating particles as a function of time. We found that the observed spectral shape of photons can be reproduced by a charged lepton population with a flat low-energy spectral index and a log-parabola energy distribution $f(\gamma)$. The required density, $N=1000\, \mathrm{cm^{-3}}$, is typical of the immediate environment surrounding an AGN, and the blob size is consistent with a variability timescale on the order of $1$ day in the observed frame. The number of emitting particles per unit volume $N$ is given by: 

\begin{equation}
    N\propto \int^{\gamma_{max}}_{\gamma_{min}} f(\gamma) d\gamma 
\end{equation} 

\noindent where we adopted a constant value of $\gamma_{min} = 2$ for the minimum Lorentz factor of the emitting particles energy distribution. The spectral distribution of particles is described by a log-parabola:

\begin{equation}
    f(\gamma)=\Big(\frac{\gamma}{\gamma_0}\Big)^{-(s+r Log[\gamma/\gamma_0])}
\end{equation}

\noindent where $\gamma_0$ is the reference energy, $r$ the curvature, $s$ the spectral index.

\begin{table}[t]
\centering
\caption{List of JetSet model parameter values that evolve over time. The reported parameters include the time, the maximum particle Lorentz factor for the energy distribution of the emitting particles ($\gamma_\mathrm{max}$), the curvature of log-parabola ($r$), the spectral index of log-parabola ($s$), and the reference energy of log-parabola ($\gamma_0$). The minimum particle Lorentz factor $\gamma_{\mathrm{min}}$ is assumed to be always $2.0$.} \label{tab:ModelPars}
\begin{tabular*}{\linewidth}{l@{\extracolsep{\fill}}ccccc}
\hline
\hline
Obs. Date & $\gamma_\mathrm{min}$ & $\gamma_\mathrm{max}$ & $r$ & $s$ & $\gamma_0$ \\
\hline
2023-05-03 & 2.0 & $5\times 10^5$ & $1.05$ & $1$ & 580 \\
2023-05-04 & 2.0  & $4\times 10^5$ & $1.05$ & $1.05$ & $575$ \\
2023-05-06 & 2.0  & $4\times 10^5$ & $1.07$ & $1.07$ & $540$ \\
2023-05-08 & 2.0  & $3\times 10^5$ & $1.15$ & $1.15$ & $540$ \\
2023-05-09 & 2.0  & $3\times 10^5$ & $1.15$ & $1.2$ & $530$ \\
\hline
\end{tabular*}
\end{table}

A comparison of the adopted model with JetSeT \citep{Massaro06,Tramacere09,Tramacere11} and the broadband observed SED is illustrated in Fig.~\ref{fig:sedmulti}. Interestingly, the model that fits the data of 2023-05-03 also provides an excellent baseline for fitting the SED observed in the following days,  except for some tension on the $\gamma$-ray emission seen on 2023-05-08, just by reducing the energy of the radiating particles $\gamma_{max}$, applying a softer spectral index $s$, and a more pronounced spectral curvature $r$. Notably, we observe that the IC component in the later days enters the hard X-ray domain, in agreement with the non decreasing trend seen only for this band in Fig.~\ref{plot:totallc}. This supports the idea that the  peak of the flare can be interpreted as a single-zone outburst, where SSC radiation played a dominant role, likely due to the acceleration of a distribution of charged particles within the jet, which subsequently cooled down through radiative losses. Unfortunately, no radio data were collected during the flaring activity presented in this paper,  therefore we can neither verify nor exclude that other processes were involved, at least on the long term. For completeness, Fig.~\ref{fig:sedmulti} also shows archival, non-simultaneous data extracted from the SSDC SED Builder and NED archives, covering facilities from radio to gamma-rays, to emphasize the magnitude of the peak activity with respect to the historical properties of 3C~216.

The tension observed between the first gamma-ray upper limit and the model of 2023-05-08 can be attributed to limited statistics, which may lead to an inappropriate representation of the spectral index in the estimate of the lowest energy $\gamma$-ray UL. Additionally, we need to take into account the non strictly simultaneous nature of the data, since our observations in the optical, UV, and X-ray bands are representative of less than $1\,$hr of exposure each, while the gamma-ray data represent the source visibility integrated over a full day. For the last daily bin on May 9, despite marginal agreement between the model predictions and the observations, it becomes more challenging to achieve a satisfactory fit within the framework of our simple energy-loss single-zone SSC model. This may be due to the flaring event losing power and  being no longer dominant over other emission regions within the source,  thus leading to a break-down of the single-zone emission assumption at later times of the event.

\begin{figure}[t]
    \centering    \includegraphics[width=1.1\linewidth]{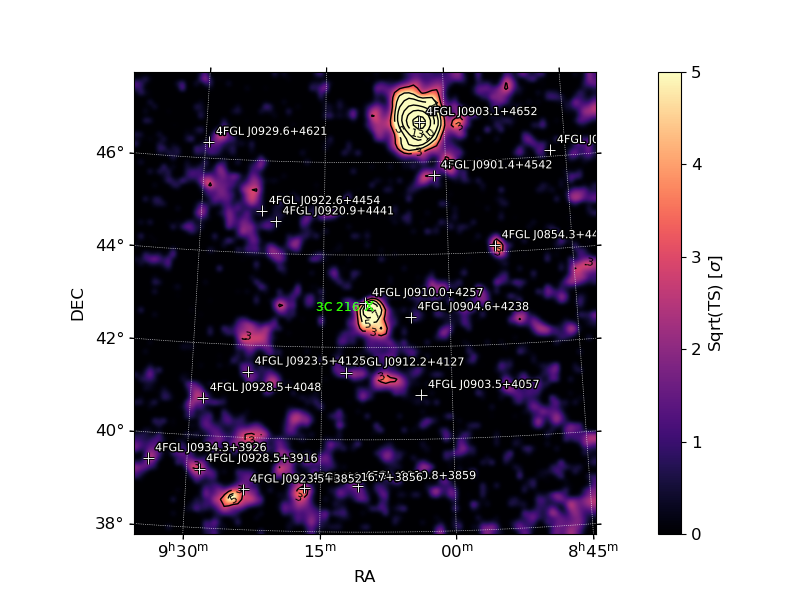}
    \caption{\textit{Fermi}-LAT $\sqrt{TS}$ map between $100$ MeV and $300$ GeV of the region around 4FGL J0910.0+4257. In green the 3C~216 position from~\citet{2005AJ....129.1163P}.}
    \label{plot:tsmap}
\end{figure}
\section{Conclusions}
\label{sec:conclusions}
The simultaneous flaring activity observed in optical, UV, X-ray and gamma-rays between 2023-05-01 and 2023-05-09 provides a confirmation of the identification of the $\gamma$-ray source 4FGL~J0910.0+4257 with 3C~216. The TS map of 4FGL J0910.0+4257 in the \emph{total} period is shown in Fig.~\ref{plot:tsmap}. We marked in green the position of 3C~216, as reported by \citet{2005AJ....129.1163P}. 
Our results also suggest that the SSC process can well explain the production of $\gamma$-ray outbursts from this type of radio sources, thus being more universally applicable across different AGN classes than previously thought. In VLBA archive\footnote{\url{https://science.nrao.edu/facilities/vlba/data-archive}}, there are data to study the morphology of the jet close to the central engine of the source. The results of our investigation could serve as foundation for future radio Target of Opportunity proposals if further high-energy flaring activity occurs.

Due to its overall steep radio spectrum and relatively compact morphology, 3C~216 is classified as a CSS source.  In the \textit{youth scenario} proposed by \citet{1995A&A...302..317F}, these sources owe their compactness to a young age and represent an early evolutionary stage towards a fully developed radio galaxy size.  CSS sources, and their even more compact (and younger) siblings (GHz Peaked Spectrum, or GPS, sources, \citealt{1991ApJ...380...66O}), were predicted to be good candidates for GeV emission, as they combine on sub-galactic scales both the presence of recently injected relativistic particles and abundant photon fields from the central regions of their hosts \citep{2008ApJ...680..911S}. However, only a handful of CSS/GPS sources have been detected individually \citep{2016ApJ...821L..31M,2020A&A...635A.185P} and even a stacking analysis has not revealed a collective signal from this population \citep{2021MNRAS.507.4564P}.

Except for a few outstanding sources detected because of their extreme proximity, a significant contribution from Doppler beamed components, such as relativistic jets, is required for CSS and GPS sources to be detected in $\gamma$ rays. This is suggested by elements such as the position in the luminosity-photon index diagram, the identification with quasar hosts, and the presence of variability.  While \citet{2021MNRAS.507.4564P} already pointed out all these characteristics for 3C~216, the flare analyzed here reveals a level of flux variability significantly more extreme than anything observed before in this source.

As far as the radio properties are concerned, noteworthy features include an upturn in the integrated spectrum around a few GHz \citep{1995AJ....110..522T}, the presence of a compact central component with flat spectrum and a strong bend, and the detection of superluminal motions on parsec scales \citep{1993A&A...271...65V,2000PASJ...52..983P}. These factors clearly indicate that along with the non-relativistic steep spectrum lobes, which may be seen in projection, the source has a blazar core seen under a small viewing angle.
\section*{Acknowledgments}
The \textit{Fermi} LAT Collaboration acknowledges generous ongoing support from a number of agencies and institutes that have supported both the development and the operation of the LAT as well as scientific data analysis. These include the National Aeronautics and Space Administration and the Department of Energy in the United States, the Commissariat \`a l'Energie Atomique and the Centre National de la Recherche Scientifique / Institut National de Physique Nucl\'eaire et de Physique des Particules in France, the Agenzia Spaziale Italiana and the Istituto Nazionale di Fisica Nucleare in Italy, the Ministry of Education, Culture, Sports, Science and Technology (MEXT), High Energy Accelerator Research Organization (KEK) and Japan Aerospace Exploration Agency (JAXA) in Japan, and the K.~A.~Wallenberg Foundation, the Swedish Research Council and the Swedish National Space Board in Sweden.

INFN and ASI personnel performed in part under ASI-INFN Agreements No. 2021-43-HH.0. 

The NASA \textit{Swift} gamma-ray burst explorer is a MIDEX Gamma Ray Burst mission led by NASA with participation of Italy and the UK. This research has made use of  the Smithsonian/NASA's ADS bibliographic database.  This research has made use of the NASA/IPAC NED database (JPL CalTech and NASA, USA). This research has made use of the archives and services of the Space Science Data Center (SSDC), a facility of the Italian Space Agency (ASI Headquarter, Rome, Italy). This research has made use of the XRT Data Analysis Software (XRTDAS) developed under the responsibility of the SSDC.

Supported by Italian Research Center on High Performance Computing Big Data and Quantum Computing (ICSC), project funded by European Union - NextGenerationEU - and National Recovery and Resilience Plan (NRRP) - Mission 4 Component 2 within the activities of Spoke 3 (Astrophysics and Cosmos Observations).

We also gratefully acknowledge Giacomo Principe, Filippo D'Ammando, Melissa Pesce-Rollins, Deirdre Horan, and anonymus external referees whose comments and suggestions contributed to shaping this work.


Facilities: \textit{Fermi Gamma-ray Space Telescope} --- \textit{LAT}  --- \textit{Swift}  --- \textit{XRT} --- \textit{UVOT} ---



\bibliographystyle{apj}

{}

\end{document}